\pdfoutput=1

\documentclass[final,3p,times,authoryear]{elsarticle}

\usepackage{amssymb}
\usepackage{amsmath}
\usepackage{float}
\usepackage{multirow}

\journal{Remote Sensing of Environment}

\begin{document}

\begin{frontmatter}



\title{Estimation of boreal forest biomass from ICESat-2 data using hierarchical hybrid inference}


\author[uef]{Petri Varvia\corref{cor1}}
\cortext[cor1]{Corresponding author}
\ead{petri.varvia@uef.fi}
\author[nmbu]{Svetlana Saarela}
\author[uef]{Matti Maltamo}
\author[luke]{Petteri Packalen}
\author[nmbu]{Terje Gobakken}
\author[nmbu]{Erik N{\ae}sset}
\author[slu]{G\"{o}ran St\aa hl}
\author[uef]{Lauri Korhonen}

\affiliation[uef]{organization={School of Forest Sciences, University of Eastern Finland },
            addressline={P.O. Box 111}, 
            city={Joensuu},
            postcode={FI-80101}, 
            country={Finland}}
\affiliation[nmbu]{organization={Faculty of Environmental Sciences and Natural Resource Management, Norwegian University of Life Sciences},
            addressline={P.O. Box 5003, NMBU}, 
            city={\r{A}s},
            postcode={NO-1432}, 
            country={Norway}}
            
\affiliation[luke]{organization={Natural Resources Institute Finland},
            addressline={Latokartanonkaari 9}, 
            city={Helsinki},
            postcode={FI-00790}, 
            country={Finland}}
\affiliation[slu]{organization={Faculty of Forest Sciences, Swedish University of Agricultural Sciences},
            addressline={SLU Skogsmarksgrand 17}, 
            city={Ume\r{a}},
            postcode={SE-90183}, 
            country={Sweden}}
           
\begin{abstract}
The ICESat-2, launched in 2018, carries the ATLAS instrument, which is a photon-counting spaceborne lidar that provides strip samples over the terrain. While primarily designed for snow and ice monitoring, there has been a great interest in using ICESat-2 to predict forest above-ground biomass density (AGBD). As ICESat-2 is on a polar orbit, it provides good spatial coverage of boreal forests.

The aim of this study is to evaluate the estimation of mean AGBD from ICESat-2 data using a hierarchical modeling approach combined with rigorous statistical inference. We propose a hierarchical hybrid inference approach for uncertainty quantification of the AGBD estimated from ICESat-2 lidar strips. Our approach models the errors coming from the multiple modeling steps, including the allometric models used for predicting tree-level AGB. For testing the procedure, we have data from two adjacent study sites, denoted Valtimo and Nurmes, of which Valtimo site is used for model training and Nurmes for validation.

The ICESat-2 estimated mean AGBD in the Nurmes validation area was 63.2$\pm$1.9 Mg/ha (relative standard error of 2.9\%). The local reference hierarchical model-based estimate obtained from wall-to-wall airborne lidar data was 63.9$\pm$0.6 Mg/ha (relative standard error of  1.0\%). The reference estimate was within the 95\% confidence interval of the ICESat-2 hierarchical hybrid estimate. The small standard errors indicate that the proposed method is useful for AGBD assessment. However, some sources of error were not accounted for in the study and thus the real uncertainties are probably slightly larger than those reported. 

\end{abstract}


\begin{keyword}
ICESat-2 \sep above-ground biomass \sep boreal forest \sep inference \sep lidar


\end{keyword}

\end{frontmatter}

\section{Introduction}
Satellite lidars have potential to improve the accuracy of global above-ground biomass (AGB) surveys by providing information on the forest height \citep{duncanson2019}. Research into using spaceborne lidar data in AGB estimation started with the first ICESat mission \citep[e.g.][]{lefsky2005,boudreau2008,nelson2017}. ICESat has since been followed by GEDI, a dedicated forest observation mission on the International Space Station \citep{dubayah2020}, and ICESat-2, launched in 2018 \citep{markus2017}. 

The ICESat-2 carries the ATLAS (Advanced Topographic Laser Altimeter System) instrument which is a profiling photon counting lidar operating at green wavelength (532 nm). ICESat-2 data consist of parallel ground tracks produced by the three pairs of strong and weak beams, which have a power ratio of 4:1 \citep{neumann2019}. While primarily designed for snow and ice monitoring, ICESat-2 has the advantage of providing a good coverage of the boreal zone, the northern parts of which are not covered by GEDI.

The current spaceborne lidar sensors have a limitation that the measurements consist of either discrete footprints or, in the case of ICESat-2, discrete height profiles. As it is unlikely that the discrete footprints or profiles overlap with the ground sites with AGB field measurements, the construction of regression models that link AGB with the satellite measurements is more complicated than, for example, in optical satellite imagery provided as spatially continuous data. The current practice is to use airborne laser scanning (ALS) to bridge the gap between reference field measurements and the satellite measurements \citep[e.g.][]{wulder2012}, either by constructing an intermediate proxy model \citep[e.g.][]{margolis2015,holm2017,narine2020,varvia2022,guerra2022} or by simulating satellite lidar measurements from the ALS data \citep{narine2019b,duncanson2022}. It is also possible to measure field plots directly at the space lidar footprint or track locations \citep{lefsky2005,nelson2009,song2022}, although in practice it is often not feasible due to e.g. poor accessibility of the footprint locations.

In addition to producing estimates of AGB or its areal density (AGBD), it is important to quantify the uncertainty of the estimated values, for example, by variance estimation. However, estimating the variance of the estimated AGB is complicated by the hierarchical modeling approach and the methodology has matured only relatively recently. As the response variables of the spaceborne lidar model are not coming from field measurements, but are predictions from the linking proxy model, they have an associated uncertainty. In the earliest studies, such as \citet{nelson2009}, the uncertainty from this model hierarchy was omitted due to intractability. 

The first study to account for multiple modeling steps in space lidar application was \citet{holm2017}, which used the so-called hybrid inference approach \citep{stahl2011} and reformulation of the hierarchical modeling to a more tractable form. In hybrid inference, the model predictions at the satellite lidar footprint or track level are treated in a similar way as observations in traditional design-based inference, such as measured sample plots in a forest inventory. As the observations are model predictions with associated uncertainty, a hybrid estimate combines the variance coming from the sample design with the propagated variances coming from model uncertainties. Hybrid inference has been used in several previous space lidar studies \citep[e.g.][]{healey2012,neigh2013,margolis2015,nelson2017,patterson2019}.

A parallel development to variance estimation in the case of hierarchical modeling was the so-called hierarchical model-based (HMB) approach \citep{saarela2016,saarela2020}, which was originally applied to a scenario where a proxy ALS model is used to link field plot data to wall-to-wall satellite imagery. HMB has since been applied also to satellite lidar applications, such as GEDI \citep{saarela2018,gedicomp}. 

The aim of this study is to evaluate the estimation of the mean AGBD and its variance using a hierarchical modeling procedure with ICESat-2, Sentinel-2, ALS and field data. We propose a hierarchical hybrid inference approach that combines error propagation through the model hierarchy in HMB with the hybrid inference approach \citep{methods}. The estimation approach is similar to what has been previously done with GEDI data \citep[e.g.][]{patterson2019,dubayah2022}, but has been modified to work with ICESat-2 and includes the uncertainty of the allometric models used to produce field-plot AGBD values. To the authors' knowledge, this is the first study where hybrid inference is used with ICESat-2 data. 

\section{Materials and methods}

\subsection{Study sites and field measurements}
\begin{figure}[h]
	\centering
	\includegraphics[width=70mm]{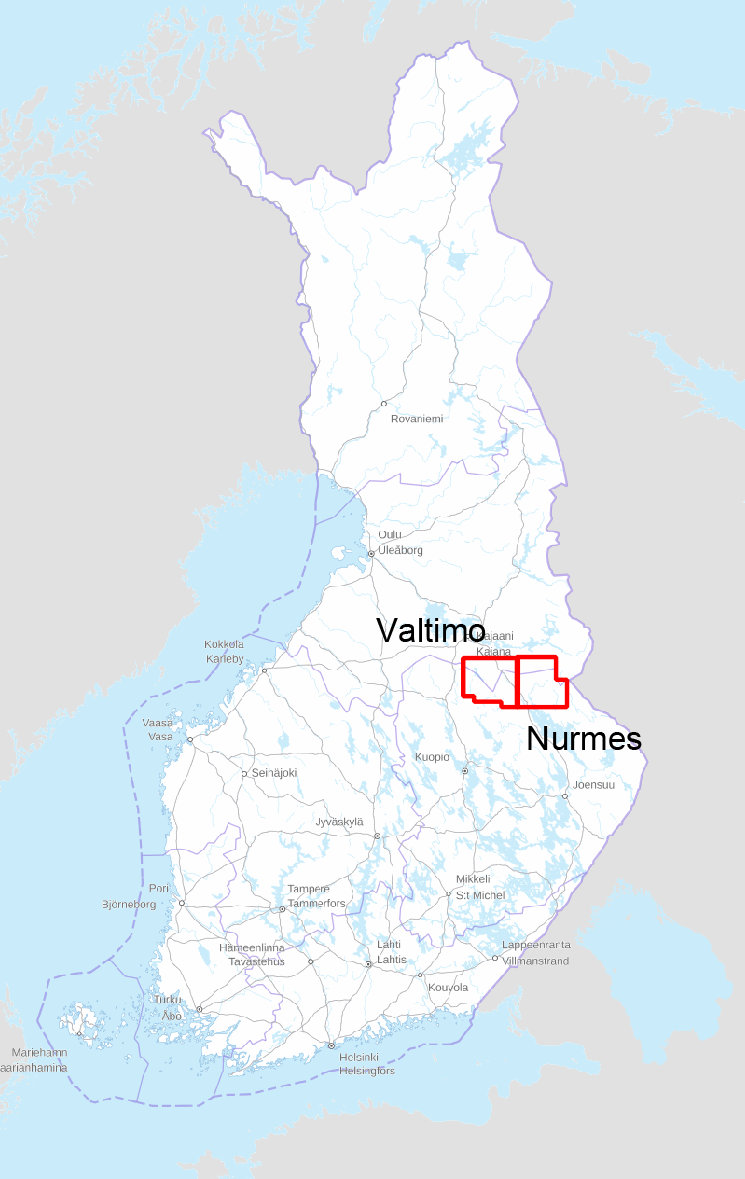}
	\caption{Location of the study sites Valtimo (N 63$^\circ$46$^\prime$ E 28$^\circ$13$^\prime$) and Nurmes (N 63$^\circ$46$^\prime$ E 29$^\circ$37$^\prime$) in Finland.}
	\label{fig:sites}
\end{figure}

The two adjacent study areas are located near Valtimo (N 63$^\circ$46$^\prime$ E 28$^\circ$13$^\prime$) and Nurmes, Finland (N 63$^\circ$46$^\prime$ E 29$^\circ$37$^\prime$). (Figure \ref{fig:sites}) Both consist of similar boreal forests, dominated by Scots pine (\textit{Pinus sylvestris} L.), with a minority of Norway spruce (\textit{Picea abies} (L.) Karst) and birches (\textit{Betula} spp.). The Valtimo site is approximately 60 $\times$ 50 km in size and the Nurmes site is 50 $\times$ 50 km. We used sample plots measured by The Finnish Forest Centre as a part of ALS-based forest management inventories in the summers 2019 and 2020 in Valtimo and Nurmes, respectively.

The field data included circular plots with radius of either 5.64 m, 9.00 m, or 12.62 m depending on the forest maturity. In total, there are 797 field plots in the Valtimo area and 891 plots in the Nurmes area. At each plot, diameter at breast height (DBH) was measured for each tree with DBH $\geq$ 5 cm. The height of a sample tree of each species was measured on each plot and a calibrated height model \citep{eerikainen2009} was used to predict the height for the rest of the trees. A summary of the field-plot data is presented in Table \ref{tab:plots}.

\begin{table}[h]
\centering
\caption{Summary of the field plot data. Height is the plot average. SD is standard deviation.}
\begin{tabular}{lcccc} \hline\noalign{\vspace{3pt}}
 & Mean & SD & Min & Max  \\ \hline
Valtimo & & & & \\
Height [m] & 10.2 & 5.5 & 0 & 25.0 \\
DBH [cm] & 10.9 & 6.3  & 0 & 35.0\\
AGBD [Mg/ha] & 62.6 & 49.5 & 0 & 298.8 \\ \hline
Nurmes & & & &  \\
Height [m] & 10.7 & 5.4 & 0 & 23.8\\
DBH [cm] & 11.9 & 6.7 & 0 & 31.8\\
AGBD [Mg/ha] & 66.5 & 57.4 & 0 & 282.8 \\ \hline
\end{tabular}
\label{tab:plots}
\end{table}

\subsection{ALS data}
The ALS data in the Valtimo area were collected between June 7th and July 9th 2019 using a Leica ALS 80 HP scanner. The flying altitude was 1700 m above ground level, resulting in a nominal pulse density of 5 pts/m$^2$ and a footprint diameter of 39 cm. In the Valtimo area, the publicly available data were used, which were resampled from 5 pts/m$^2$ to 0.5 pts/m$^2$ before distribution. The ALS data in the Nurmes area were collected between June 17th and June 22nd 2020 using a Riegl VQ-1560i scanner at a flying altitude of 2100 m. In the Nurmes area the original point cloud with $>$5 pts/m$^2$ was used.

The ALS processing was done identically for both sites. The ALS echoes were height normalized with respect to ground using LAStools \citep{lastools}. For each plot, canopy metrics were computed using ``first-of-many'' plus ``only'' echoes, and ``last-of-many'' plus ``only'' echoes, producing two sets of metrics. The metrics included mean and maximum heights, standard deviation of heights, height percentiles $p_5,p_{10},p_{20},\dots,$ $p_{90},p_{95},p_{99}$, canopy density percentiles $b_5,b_{10},b_{20},\dots,b_{90},b_{95}$, canopy cover, and the mean and standard deviation of intensities.

\subsection{Sentinel-2 data}
For the Valtimo site, a cloud-free Sentinel-2 image was available from June 14th 2019. For the Nurmes site, a cloud-free Sentinel-2 composite was constructed from images captured on June 16th, July 16th, and July 18th 2020. Atmospheric correction of the Sentinel-2 images was done using Sen2Cor \citep{sen2cor}, after which the atmospheric bands (bands 1, 9, and 10)  were omitted. The images were then calibrated using histogram matching before compositing. The pixel values were used as predictors in the proxy AGBD models, in addition to several common spectral vegetation indices calculated from the images.

\subsection{ICESat-2 data}
The ATL03 \citep{atl03} and ATL08 \citep{atl08} data for the Valtimo site covered the period from October 2018 to December 2019. For the Nurmes site, ATL03 and ATL08 data captured during the year 2020 were used. Version 4 of the data products were used for both sites. 

The ICESat-2 data were processed following \citet{varvia2022}. First, the ICESat-2 tracks were split into 90 m $\times$ 15 m segments, centered on the locations of the ATL08 product. Each 90 m segment was further divided into six 15 m $\times$ 15 m subcells, which were used for the prediction of proxy AGBD on the 90 m track segments. The ATL08 individual photon classifications were then matched with the photon locations from ATL03 product. Photons classified as noise were discarded. 

The classified photons were then clipped to the 90 m track segments. Using the photons classified as ground, the above-ground height was computed for each photon. Several height metrics were then calculated, and similar to the ALS metrics detailed above, they included the number of photons (canopy only ($n_{\mathrm{c}}$) and total ($n_{\mathrm{all}}$)), mean photon height, standard deviation, maximum, height percentiles $p_5,p_{10},p_{20},\dots,p_{90},p_{95},p_{99}$, canopy density percentiles $b_5,b_{10},b_{20},\dots,b_{90},b_{95}$, and mean square height ($\mathrm{qav}$). 

Poor quality segments were omitted if they did not meet the criteria of at least 100 classified photons and a fraction of high confidence photons (\verb|signal_conf_ph| in ATL03) being at least 60\%. In addition, a polygonal forest mask produced by the Finnish Forest Centre \citep{forestmask} was used to discard ICESat-2 segments in certain non-forested areas, such as agricultural fields, water, roads, and built-up areas.

\begin{figure}[tb]
	\centering
	\includegraphics[width=140mm]{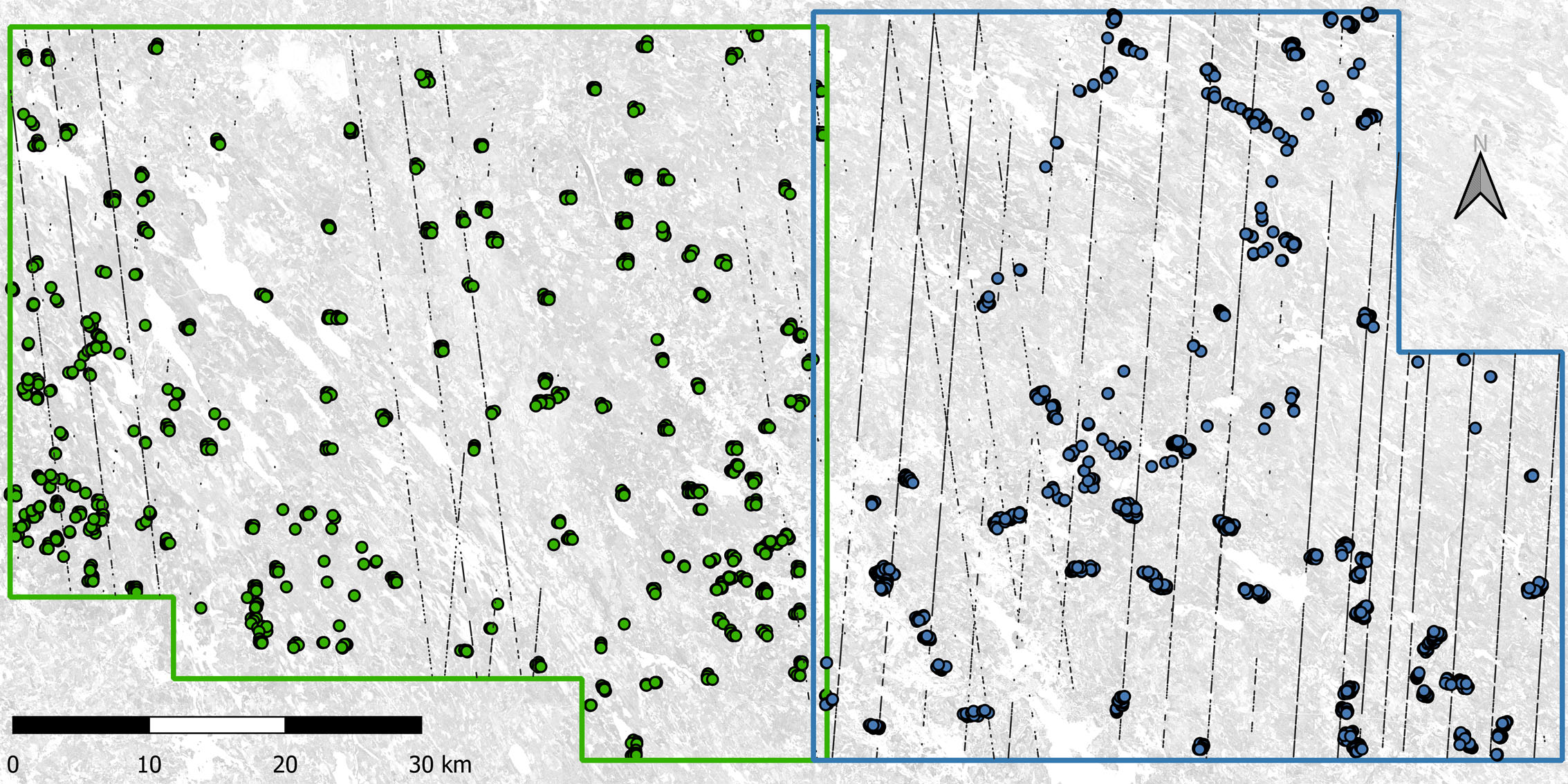}
	\caption{The locations of 90 m ICESat-2 segments in Valtimo (left, green) and Nurmes (right, blue) superimposed on a canopy height map (Finnish Forest Centre). Field plot locations shown with circles.}
	\label{fig:segs}
\end{figure}

We used only strong beam data captured outside the snowy season during daytime. While night data would be preferable due to the absence of solar noise, not enough snowless strong beam night data were available from the Nurmes site (only 60 segments). The use of daytime data thus represents a compromise between expected performance and data availability. For the Valtimo area there was a total of 1721 valid 90 m segments and 5760 segments in Nurmes (Figure \ref{fig:segs}).

\subsection{Methods}
The process of estimating AGBD using ICESat-2 data follows hierarchical modeling with three steps: 1) deriving field-plot AGBD from allometric AGB models and measured trees, 2) model based on ALS and Sentinel-2 data for predicting proxy AGBD on ICESat-2 tracks, and 3) ICESat-2 model for predicting AGBD from ICESat-2 height metrics. The model chain is trained using the data from Valtimo area and the fitted ICESat-2 model is then used to predict AGBD with the Nurmes data. 

We used the species-specific biomass models by \citet{repola08,repola09} to predict allometric AGB for each measured tree on the field plots using the calipered DBH and predicted tree height. The field-plot AGBD was then calculated by summing up the tree-level AGBs of each plot and scaled to per-hectare level (\ref{sec:allo}) .

A quadratic model with four variables was then fitted between the field-plot AGBD, and ALS and Sentinel-2 metrics using generalized nonlinear least squares. The four metrics in the model were chosen using a simulated annealing based variable selection routine \citep{packalen2012}. The ALS and Sentinel-2 model was then used to predict proxy AGBD on the 15 m $\times$ 15 m subcells constructed on the ICESat-2 tracks. The subcell AGBDs were averaged to calculate proxy AGBD values for the 90 m track segments used in the ICESat-2 modeling (\ref{sec:als}).

In the final modeling step, a quadratic model with four variables was fitted between the 90 m proxy AGBD and ICESat-2 metrics; the four metrics were again chosen using simulated annealing. This model was then used to predict AGBD on the Nurmes ICESat-2 track segments. (\ref{sec:i2})

The AGBD predictions on the Nurmes ICESat-2 track segments were used to calculate a hierarchical hybrid estimate for the mean and variance of AGBD in the Nurmes area. The predictions on the track segments were modeled as a clustered random sample \citep{stahl2011}, where each ICESat-2 track was a cluster, similar to what was previously done by \citet{dubayah2022} with GEDI data. As the tracks can cross and overlap, the design is considered as sampling with replacement. The estimate for mean AGBD is
\begin{equation}
\hat{\mu}_{\mathrm{I2}} = \frac{\sum_{i=1}^{n_{\mathrm{track}}}\widehat{\mathrm{AGBD}}_{\mathrm{I2,sum}}^{(i)}}{\sum_{i=1}^{n_{\mathrm{track}}}n_{\mathrm{seg}}^{(i)}},
\end{equation}
where $\widehat{\mathrm{AGBD}}_{\mathrm{I2,sum}}^{(i)}$ is the summed up predicted AGBD of the $i$'th ICESat-2 track, $n_{\mathrm{seg}}^{(i)}$ is the number of 90 m segments in the $i$'th track, and $n_{\mathrm{track}}$ is the number of tracks.

\begin{figure}[H]
	\centering
	\includegraphics[width=70mm]{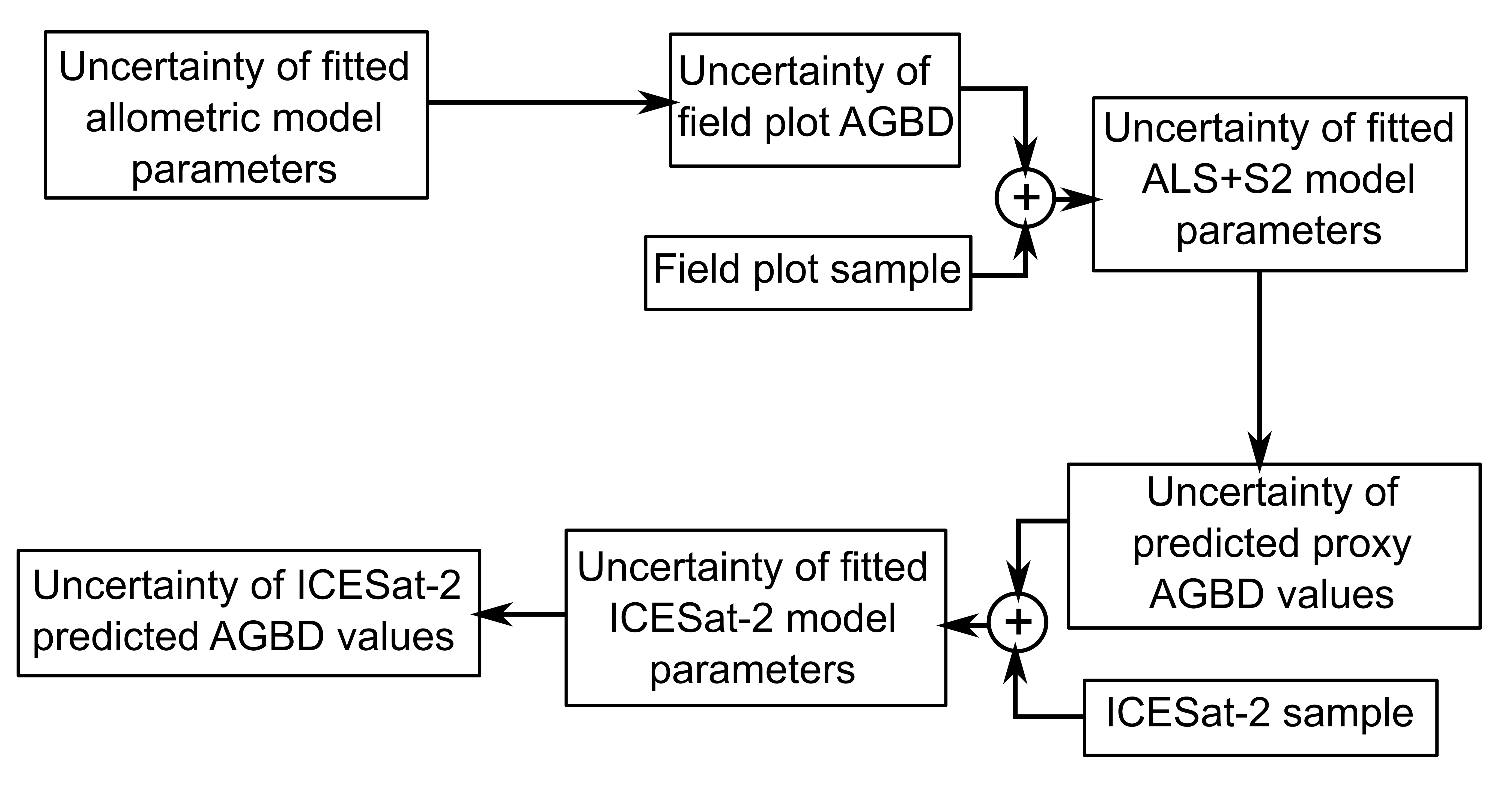}
	\caption{Schematic of the error propagation.}
	\label{fig:hier}
\end{figure}

The hybrid estimator for AGBD variance consists of two parts. First is the design-based sampling variability under the assumed design
\begin{equation}
\widehat{\mathrm{Var}}_D(\hat{\mu}_{\mathrm{I2}})=\frac{1}{\bar{n}^2_{\mathrm{seg}}}\frac{\sum_{i=1}^{n_{\mathrm{track}}}(\widehat{\mathrm{AGBD}}_{\mathrm{I2,sum}}^{(i)}-\hat{\mu}_{AGBD}n_{\mathrm{seg}}^{(i)})^2}{n_{\mathrm{track}}(n_{\mathrm{track}}-1)},
\end{equation}
where $\bar{n}_{\mathrm{seg}}$ is the average number of segments per track. The second part is the model-based uncertainty of the predicted AGBDs, which is
\begin{equation}
\widehat{\mathrm{Var}}_M(\hat{\mu}_{\mathrm{I2}})=\frac{1}{n_{\mathrm{tot}}^2}\boldsymbol{1}^T\widehat{\mathbf{C}}_{\mathrm{I2}}\boldsymbol{1},
\end{equation}
where $n_{\mathrm{tot}}$ is the total number of ICESat-2 segments and $\boldsymbol{1}$ is a vector of ones. The covariance matrix $\widehat{\mathbf{C}}_{\mathrm{I2}}$ is the hierachical model-based covariance of the ICESat-2 predictions. For the derivation of $\widehat{\mathbf{C}}_{\mathrm{I2}}$, see Appendix. Schematic of the error progation in the model-based part is shown in Figure \ref{fig:hier}. Finally, the estimated variance of the mean AGBD is
\begin{equation}
\widehat{\mathrm{Var}}(\hat{\mu}_{\mathrm{I2}})  = \widehat{\mathrm{Var}}_D(\hat{\mu}_{\mathrm{I2}}) + \widehat{\mathrm{Var}}_M(\hat{\mu}_{\mathrm{I2}}).
\end{equation}
Additional details and derivation of the hierarchical hybrid approach are presented in the companion article by \citet{methods}.

\subsubsection{Reference estimate}
Hierarchical model-based (HMB) estimate calculated using an independent local data set was used as a reference estimate of mean AGBD at the Nurmes validation area (see \ref{sec:ref}). As in the ICESat-2 workflow, a quadratic model with four variables was fitted between the field-plot AGBD, and ALS and Sentinel-2 metrics in Nurmes. The model was then used to produce a 15 m $\times$ 15 m wall-to-wall AGBD raster over the area. The variance estimation procedure followed \citet{saarela2020} and included the uncertainty from the allometric models and the ALS and Sentinel-2 model fitted using the Nurmes field plots. 

In addition to comparing the estimated mean AGBD and its variance, ICESat-2 model performance was evaluated using root mean square deviation (RMSD) \eqref{rmse} and mean difference (MD) \eqref{bias}. The ICESat-2 predictions in the Nurmes area were compared to AGBD predicted on the ICESat-2 segment locations using the local ALS and Sentinel-2 model.

\begin{align}
\label{rmse}
\mathrm{RMSD}&=\sqrt{\frac{1}{n}\sum\left(\widehat{\mathrm{AGBD}}_{\mathrm{I2}}-\widehat{\mathrm{AGBD}}_{\mathrm{proxy}}\right)^2} \\
\label{bias}
\mathrm{MD}&=\frac{1}{n}\sum\left(\widehat{\mathrm{AGBD}}_{\mathrm{I2}}-\widehat{\mathrm{AGBD}}_{\mathrm{proxy}}\right)
\end{align}

\section{Results}
\subsection{AGBD models}
The ALS and Sentinel-2 AGBD models and the ICESat-2 AGBD had a similar quadratic form with four predictors chosen using a simulated annealing. The fitted ALS and Sentinel-2 proxy AGBD model in the Valtimo area was:
\begin{equation}
\label{eq:bmmodel}
\begin{split}
\widehat{\mathrm{AGBD}}_{\mathrm{proxy}}=&(8.52+0.44\;\mathrm{avg}_{f}+0.34\;\mathrm{avg}_{l}\\
&- 0.0013\;\mathrm{NIR} -0.0012\;\mathrm{SWIR1})^2.
\end{split}
\end{equation}
The model was fitted using generalized nonlinear least squares \citep{nlme}, with an estimated residual variance function of a constant plus power structure
\begin{equation}
\widehat{\mathrm{Var}}(\mathbf{e}_f)=0.61^2\left(6.23+\widehat{\mathrm{AGBD}}_{\mathrm{proxy}}^{0.69}\right)^2.
\end{equation}

Similarly, the ALS and Sentinel-2 model for the Nurmes validation area, which was used to obtain the reference HMB estimate, was
\begin{equation}
\label{eq:bmmodelnurmes}
\begin{split}
&\widehat{\mathrm{AGBD}}_{\mathrm{ref}}=(9.77-0.034b_{10,f}+0.40p_{99,l}\\
&- 0.042b_{40,l} -0.0012\;\mathrm{NIR})^2,
\end{split}
\end{equation}
and the residual variance function was
\begin{equation}
\widehat{\mathrm{Var}}(\mathbf{e}_\mathrm{ref})=0.99^2\left(2.78+\widehat{\mathrm{AGBD}}_{\mathrm{ref}}^{0.61}\right)^2.
\end{equation}

The proxy model \eqref{eq:bmmodel} was applied to predict AGBD for the ICESat-2 segments at the Valtimo study area. These predictions were used as response AGBDs in the fitting of the ICESat-2 AGBD model in the Valtimo area:
\begin{equation}
\label{eq:bmmodeli2}
\begin{split}
\widehat{\mathrm{AGBD}}_{\mathrm{I2}}=&(1.90+1.37\;\mathrm{std}+0.12\sqrt{n_{\mathrm{c}}}\\
&+0.25\sqrt{p_{40}} -0.61\sqrt{p_{80}})^2.
\end{split}
\end{equation}
Residual variance was homoscedastic for the ICESat-2 model: $\widehat{\mathrm{Var}}(\mathbf{e}_g) = 433.0 $ Mg$^2$/ha$^2$.

\begin{figure}[H]
	\centering
	\includegraphics[width=140mm]{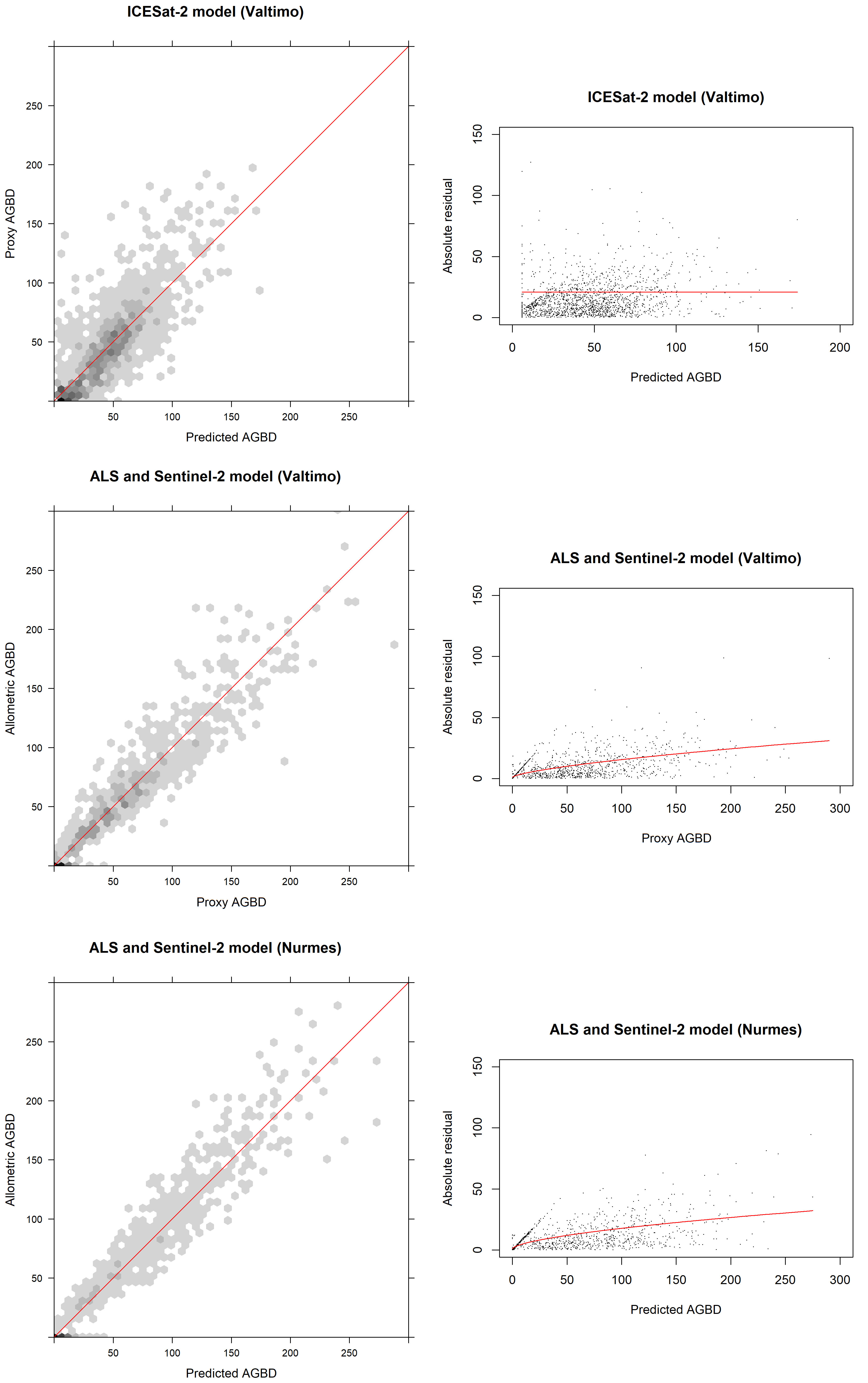}
	\caption{Scatter density plots and absolute residuals of the fitted models. Red line in the residual plots shows the standard deviation from the residual variance model used in model fitting.}
	\label{fig:fits}
\end{figure}

Scatter density plots of the fitted models are shown in Figure \ref{fig:fits}. The residual plots on the right side of Figure \ref{fig:fits} show the absolute residuals and the standard deviation of the fitted residual variance functions.

\begin{figure}[H]
	\centering
	\includegraphics[width=90mm]{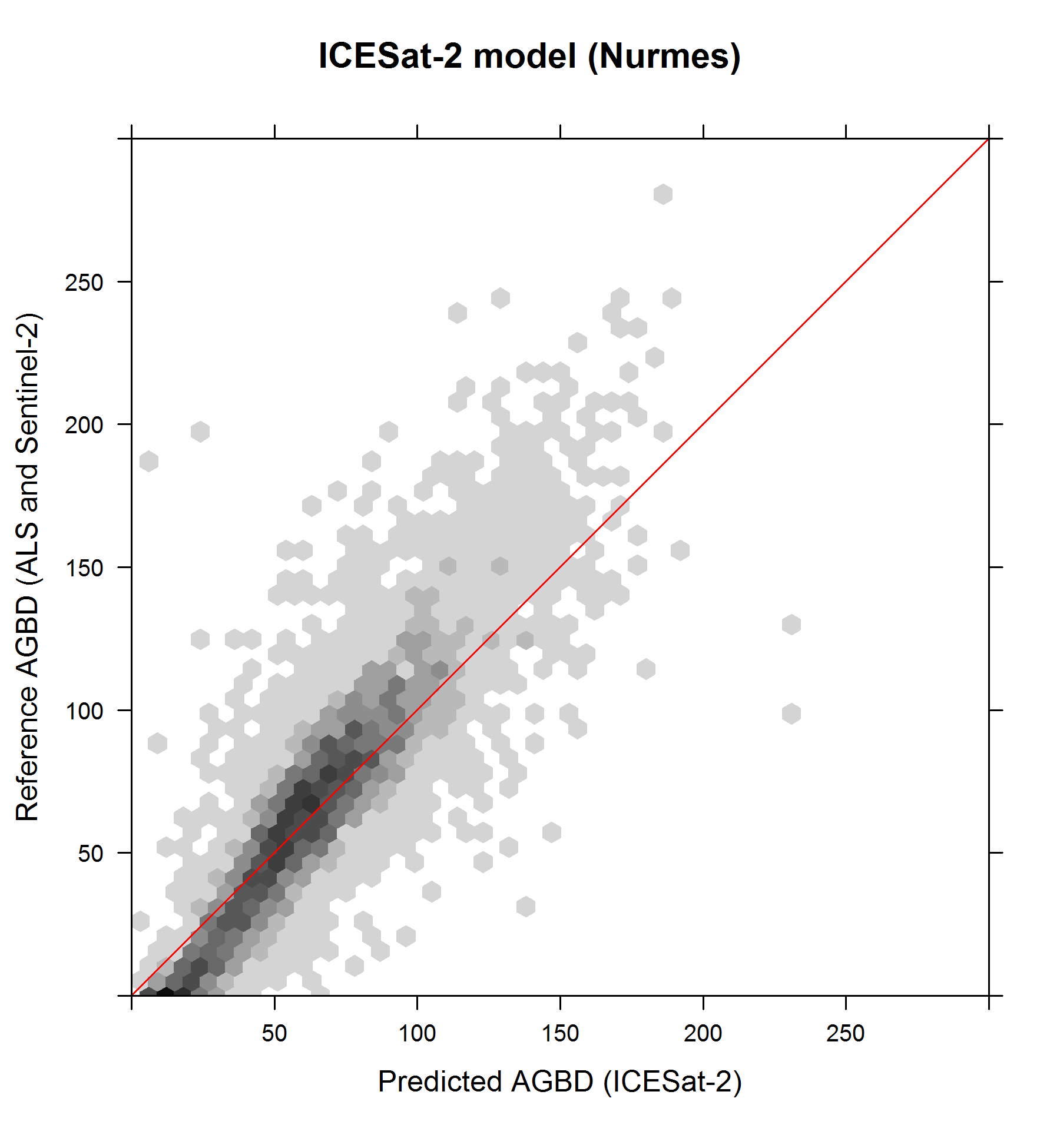}
	\caption{A comparison of AGBDs predicted by the local ALS and Sentinel-2 model vs. the ICESat-2 model fitted in Valtimo training area for the Nurmes validation area.}
	\label{fig:scatter}
\end{figure}

\begin{figure}[H]
	\centering
	\includegraphics[width=140mm]{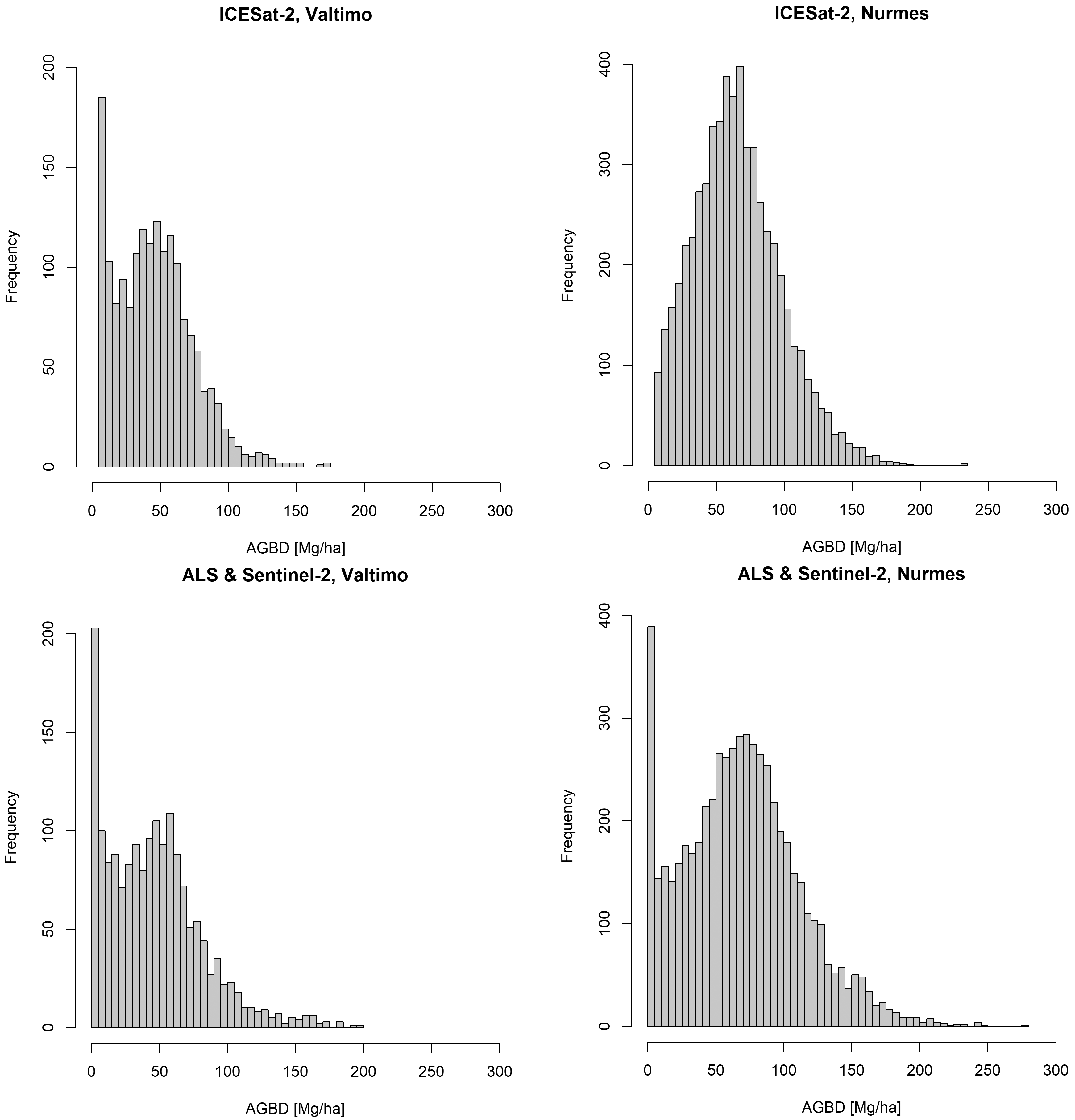}
	\caption{Histograms of the ICESat-2 model predicted AGBD and the AGBD predicted by the local ALS and Sentinel-2 models in the Valtimo and Nurmes areas.}
	\label{fig:hists}
\end{figure}

The ICESat-2 AGBD model \eqref{eq:bmmodeli2} was then applied to predict the AGBD for the ICESat-2 segments in the Nurmes validation area. The accuracy of these predictions was first compared with the predictions obtained with the local ALS and Sentinel-2 model \eqref{eq:bmmodelnurmes} at the Nurmes track locations. The ICESat-2 predictions had an RMSD of 21.3 Mg/ha (30.8\%) and an MD of -3.3 Mg/ha (-4.8\%). Scatter plot of the predictions is shown in Figure \ref{fig:scatter}. Summary of the training proxy AGBD and ICESat-2 model predictions in the Valtimo area, Nurmes area, and of the local ALS and Sentinel-2 model predictions are shown in Table \ref{tab:sum}. Histograms of the proxy AGBD and ICESat-2 predictions are shown in Figure \ref{fig:hists}.

\begin{table}[H]
\centering
\caption{Statistical summaries of the estimated AGBD values at 90 m ICESat-2 track segment level. Units are Mg/ha.}
\begin{tabular}{lcccccc} \hline\noalign{\vspace{3pt}}
Data & Min. & Q1 & Median & Mean & Q3 & Max \\
Train & & & & & & \\
Valtimo, local ALS+S2 & 0 & 17.4 & 42.7 & 46.0 & 64.9 & 199.8 \\
Valtimo, ICESat-2 & 5.9 & 23.0 & 43.9& 45.8 & 62.7 & 174.0 \\ \hline
Test & & & & & & \\
Nurmes, local ALS+S2 & 0 & 37.9 & 67.3 & 69.0 & 94.9 & 278.3 \\
Nurmes, ICESat-2 & 5.9 & 42.8 & 63.3 & 65.7 & 85.3 & 231.3 \\ \hline
\end{tabular}
\label{tab:sum}
\end{table}

\begin{figure}[H]
	\centering
	\includegraphics[width=70mm]{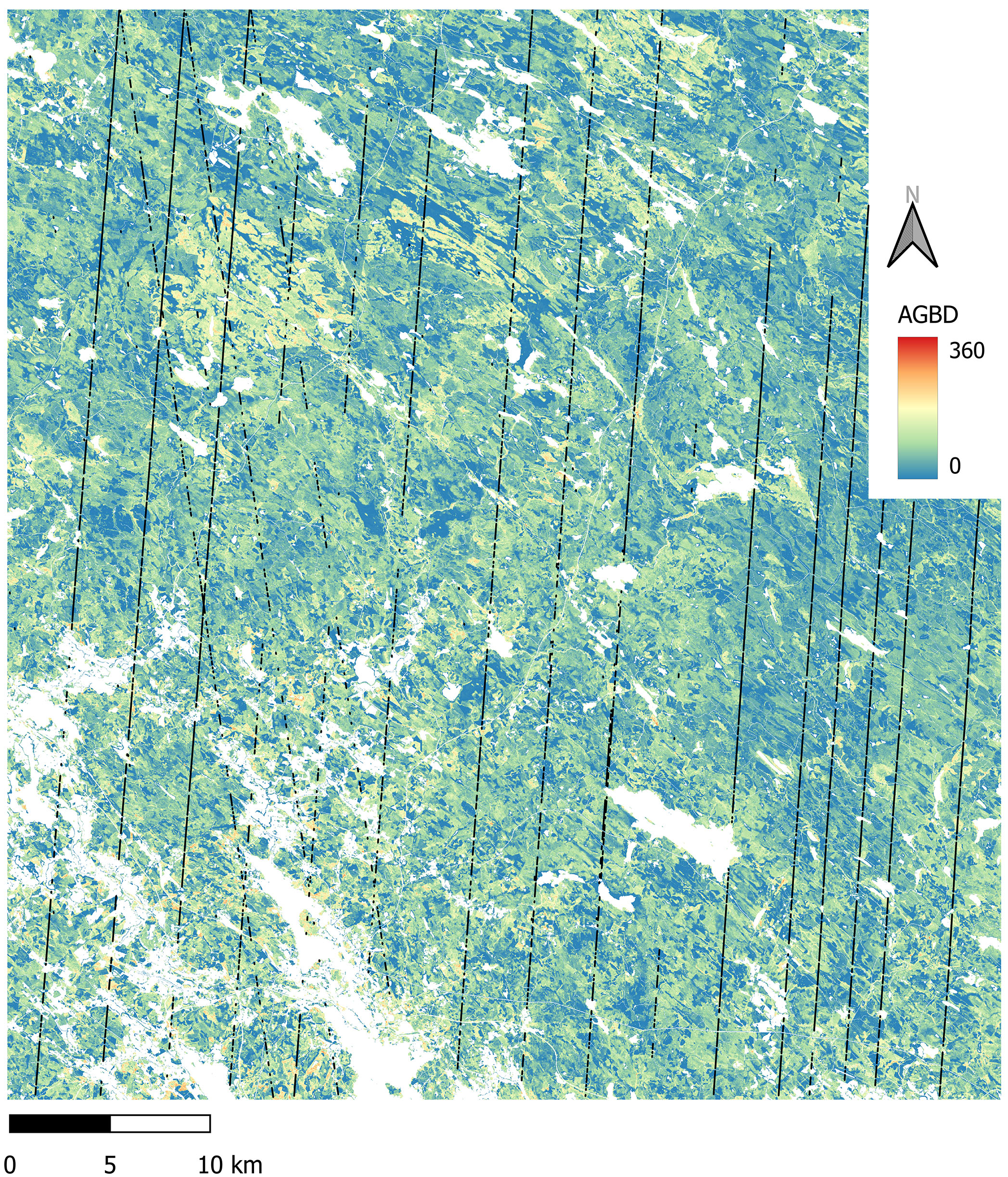}
	\caption{15 m resolution AGBD [Mg/ha] map of the Nurmes area produced using the reference ALS and Sentinel-2 model. ICESat-2 tracks shown in black.}
	\label{fig:agbmap}
\end{figure}

The 15 m resolution AGBD map produced using the local model based on ALS and Sentinel-2 data is shown in Figure \ref{fig:agbmap}. Non-forested areas were discarded. Visible features are the Hiidenportti National Park and other protected forests in the north of the area, which are visible as yellow areas of large biomass values in the map. The area also has many open mires, which show up as blue areas of small AGBD.

\subsection{Variance estimation}
The mean AGBD and its standard error over the forested parts of the Nurmes area was $65.7\pm1.9$ Mg/ha when using ICESat-2 data and hierarchical hybrid estimation (Table \ref{tab:est}). The reference hierarchical model-based estimate using local ALS and Sentinel-2 data was $63.9\pm0.6$ Mg/ha. The relative standard errors were 2.9\% and 1.0\%, respectively.

\begin{table}[H]
\centering
\caption{Estimated average AGBD in the Nurmes area, its standard error and relative standard error for the hybrid hierarchical estimation using ICESat-2 and the reference hierarchical model-based estimate using local ALS and Sentinel-2 model.}
\begin{tabular}{lccc} \hline\noalign{\vspace{3pt}}
Method & AGBD [Mg/ha] & Standard error [Mg/ha] & Relative standard error \\ \noalign{\vspace{2pt}}\hline\noalign{\vspace{3pt}}
Hierarchical hybrid & 65.7 & 1.91 & 2.9\% \\
Reference HMB & 63.9 & 0.64 & 1.0\% \\ \hline
\end{tabular}
\label{tab:est}
\end{table}

The contributions of the model hierarchy levels to the standard error in the hierarchical hybrid estimation are shown in Table \ref{tab:comp}. The topmost row shows the standard error using the full model hierarchy (1.91 Mg/ha). In the next row, the allometric part is removed, with the assumption that the allometric AGB values predicted are accurate, this decreases standard error slightly to 1.87 Mg/ha, which is 97.9\% of the total modeled standard error. Moving down, the uncertainty of the proxy AGBD model is removed; this assumes that the proxy AGBD derived from the ALS and Sentinel-2 model is accurate. This further slightly decreases the standard error to 1.81 Mg/ha (94.9\% of the total). Finally, removing the model-based uncertainty altogether, we arrive at the standard error from sample design only, which is 1.61 Mg/ha or 84.3\% of the total. By calculating the differences, the sources of uncertainty in order of decreasing magnitude are: sample design (1.61 Mg/ha, 84.3\%), ICESat-2 model (0.20 Mg/ha, 10.7 \%), proxy AGBD model (0.06 Mg/ha, 2.9\%), and allometry (0.04 Mg/ha, 2.1\%).

\begin{table}[H]
\centering
\caption{Contribution of the uncertainty components. For example, the field ``Design, I2 model'' includes only the uncertainty from the limited number of ICESat-2 segments in Nurmes and the uncertainty of the ICESat-2 model parameters when the proxy AGBD values are assumed to be accurate.}
\begin{tabular}{lcc} \hline\noalign{\vspace{3pt}}
Modeled uncertainties & \hspace{-2em}Standard error [Mg/ha] & Proportion \\ \noalign{\vspace{2pt}}\hline\noalign{\vspace{3pt}}
Design, I2, proxy AGBD, allometry\hspace{-3em} & 1.91 & 100\%\\
Design, I2, proxy AGBD & 1.87 & 97.9\%\\
Design, I2 model & 1.81 & 94.9\% \\
Sample design only & 1.61 & 84.3\%\\  \hline
\end{tabular}
\label{tab:comp}
\end{table}

\section{Discussion}
The average AGBD estimated using the ICESat-2 model transferred from a nearby area ($65.7\pm1.9$ Mg/ha) was close to the value produced using the local ALS and Sentinel model ($63.9\pm0.6$ Mg/ha). The reference estimate is within the 95\% confidence interval of the ICESat-2 estimate. However, the comparison of ICESat-2 and reference model predictions at the track segment level revealed that the ICESat-2 model had a tendency to underestimate AGBD (MD of -3.3 Mg/ha). If the estimated MD was subtracted from the ICESat-2 model predictions, the average AGBD estimate would rise to 69.0 Mg/ha. In this case, the difference to the reference estimate would be 5.1 Mg/ha. Considering that the ICESat-2 model was transferred from adjacent area and previous year, this is still a promising result.

One reason for the observed systematic error in ICESat-2 AGBD predictions was that the ICESat-2 training data from the Valtimo area have smaller AGBD values (Table \ref{tab:sum}) than the Nurmes target population, on average by 23 Mg/ha. If the estimation process was scaled up, to e.g. country level, the problem could likely be mitigated by using more training data from a larger number of locations which would better capture the full variation in the population.

The estimated relative standard error (3.5\%) of hybrid estimation is small, especially considering the complex model chain and relatively poor accuracy of the ICESat-2 model. This is partly due to using variance as an uncertainty measure: variance corresponds to the variation of the estimate around the expected value of the estimate. As seen in Table \ref{tab:sum} and Figure \ref{fig:hists}, the ICESat-2 model generally predicted biomass values that have less variation, which then also reduced the variation in the estimated mean AGBD. This seems to be an inherent property of long model chains, where each modeling step further reduces the variation \citep{methods}. This effect can be directly seen when comparing the reference HMB standard error (0.6 Mg/ha) to the model-based part in the hierarchical hybrid estimate (0.3 Mg/ha). An underlying assumption in variance as a uncertainty measure is also that the model is approximately unbiased in the target population. Based on the observed MD of the ICESat-2 predictions, the assumption likely did not hold.

Mean square error (MSE)\footnote{Not to be confused with the RMSD used previously to evaluate prediction accuracy, see e.g. \citet{gregoire2007}}  of the estimatorwould likely be a better uncertainty measure by aiming to model the discrepancy between the estimate and the true value and thus accommodate for systematic errors. However, deriving a MSE estimator for a complex model hierarchy appears to be currently intractable, for example, due to the problem of cross-correlation of the spatial autocorrelation effects at different modeling steps. Further complicating the situation, addition of spatial correlation effects can have a relatively small effect on the quantified uncertainty in some cases. For example in \citet{fortin2022} it slightly reduced the uncertainty compared to using only variance.

In this study, predicted tree heights were used in the allometric models due to unavailability of measured tree heights for all trees. This is a source of uncertainty that was omitted in the study, as it would add a further, complicating modeling step. Propagating the uncertainty from the estimated tree heights would likely require refitting of the mixed-effect models presented in \citet{eerikainen2009}. 

Interpretation of the contribution of the model components to the resulting standard error (Table \ref{tab:comp}) is complicated by two factors. First is the decrease in variation described earlier. In addition to the modeling steps, the averaging of 15 m proxy AGBDs to the 90 m segments used in the ICESat-2 model further decreases the variation coming from allometry and the ALS and Sentinel-2 model. The second factor is a problem in calculating the contributions. Previously, \citet{saarela2020} used fraction of the total variance to evaluate component contributions. We opted here to evaluate contribution using standard error by removing modeling steps one at a time and calculating the differences, primarily since standard error is in the same units as the estimate (Mg/ha) and thus could be easier to interpret. Both approaches have the limitation that the contribution of the lower modeling steps in the hierarchy cannot be evaluated directly, as they are affected by the propagation through the model chain. However, keeping these limitations in mind, the contribution of the model components to the total standard error is logical. The largest model-based contributor is the ICESat-2 model (10.7\%), which is also the least accurate model when measured by goodness of fit. The order of the contributions of allometry and ALS and Sentinel-2 model are also in line with their respective performance.

The reference HMB estimate had a considerably smaller relative standard error (1.0\%) compared to that reported by \citet{saarela2020} (7.5\%). The discrepancy seems to be mostly explained by differences in the data. The current study had a larger number of field plots, and Sentinel-2 data was used in addition to ALS, which resulted in a better performing AGBD model. The smaller positioning error of sample plots in this study ($< 1$ meter) and the placement of field plots within the forest stands (never on stand borders) may also have contributed to the reference model accuracy. In the allometric modeling, \citet{saarela2020} used separate models for trees with only measured diameter and for trees with measured diameter and height. We used models with measured diameter and estimated height for all trees. Comparatively small uncertainties in AGBD estimation have also been reported earlier by \citet{esteban2019} (1.8\%), although the study did not include allometric contribution.

\section{Conclusions}
In this study, we evaluated estimation of average AGBD using ICESat-2 data and hierarchical modeling. Uncertainty of the estimated AGBD was quantified using hierarchical hybrid inference, which combines the error propagation through the multiple modeling steps with the variance coming from the sparse spatial coverage of the ICESat-2 data.

The ICESat-2 based estimate for the Nurmes validation area was $65.7\pm1.9$ Mg/ha compared to the local reference estimate of $63.9\pm0.6$ Mg/ha. The reference estimate was within the 95\% confidence interval of the ICESat-2 based estimate. However, the interpretation was complicated by the observed presence of systematic error in the ICESat-2 AGBD predictions (-3.3 Mg/ha) at the validation area. 

While the small estimated standard error should not be interpreted in the way that the proposed ICESat-2 estimate is highly reliable, the results support the use of ICESat-2 data for AGBD estimation. In this study, the ICESat-2 model was transferred from a different year and an adjacent area with relatively good results. Further studies should consider similar estimations for larger areas where also the structure of the forest can change.

\section*{Acknowledgements}
This study was supported by the Academy of Finland (grant numbers 332707 and 352782) and the Academy of Finland Flagship Programme (Forest-Human-Machine Interplay - Building Resilience, Redefining Value Networks and Enabling Meaningful Experiences (UNITE); grant numbers 357906 and 357909).

\appendix
\section{Error propagation and variance estimation}
In the following sections, the modeling steps and the associated variance estimators are described starting from the allometric models. For derivation of the estimators, see \citet{methods}.

\subsection{Biomass allometry}
\label{sec:allo}
The species-specific biomass models by \citet{repola08,repola09} were used to predict allometric AGB for each measured tree in the field plots using the calipered DBH and predicted tree height. As the \mbox{Repola} models were fitted using log-transformation, the predicted values were corrected for bias. The predicted individual tree biomass values were then aggregated to produce plot-level AGB density $\widehat{\mathrm{AGBD}}_{\mathrm{plot}}$.

For variance estimation, we need to calculate the covariance matrix of $\widehat{\mathrm{AGBD}}_{\mathrm{plot}}$. Let us first combine the species-specific biomass models into a single model using binary species indicator variables $s_{\mathrm{pi}}$, $s_{\mathrm{sp}}$, and $s_{\mathrm{de}}$ for pine, spruce and deciduous trees, respectively:
\begin{equation}
\label{eq:allo}
\mathrm{AGB}(\hat{\boldsymbol{\alpha}},d,h,\mathbf{s}) = s_{\mathrm{pi}}\mathrm{AGB}_{\mathrm{pi}}(\hat{\boldsymbol{\alpha}}_{\mathrm{pi}},d,h)+s_{\mathrm{sp}}\mathrm{AGB}_{\mathrm{sp}}(\hat{\boldsymbol{\alpha}}_{\mathrm{sp}},d,h)+s_{\mathrm{de}}\mathrm{AGB}_{\mathrm{de}}(\hat{\boldsymbol{\alpha}}_{\mathrm{de}},d,h),
\end{equation}
where $\hat{\boldsymbol{\alpha}}= \begin{bmatrix}\hat{\boldsymbol{\alpha}}_{\mathrm{pi}},\hat{\boldsymbol{\alpha}}_{\mathrm{sp}},\hat{\boldsymbol{\alpha}}_{\mathrm{de}}\end{bmatrix}^T$, in which e.g. $\hat{\boldsymbol{\alpha}}_{\mathrm{pi}}$ is the fitted parameters for the biomass model of pine, $d$ is diameter at breast height, $h$ is tree height, and $\mathbf{s}= \begin{bmatrix}s_{\mathrm{pi}},s_{\mathrm{sp}},s_{\mathrm{de}}\end{bmatrix}^T$.

We then use Taylor approximation to calculate the covariance matrix of the tree-level AGB predictions:
\begin{equation}
\widehat{\mathbf{C}}_{\mathrm{tree}} = \mathbf{J}_{\mathrm{tree}}^T\widehat{\mathbf{C}}_{\boldsymbol{\alpha}}\mathbf{J}_{\mathrm{tree}}.
\end{equation}
where $\widehat{\mathbf{C}}_{\boldsymbol{\alpha}}=\mathrm{diag}\left(\widehat{\mathbf{C}}_{\boldsymbol{\alpha}_{pi}},\widehat{\mathbf{C}}_{\boldsymbol{\alpha}_{sp}},\widehat{\mathbf{C}}_{\boldsymbol{\alpha}_{de}}\right)$ is a block-diagonal matrix consisting of the estimated covariance matrices of the species-specific biomass models reported in \citet{stahl2014}. The matrix $\mathbf{J}_{\mathrm{tree}}$ is the Jacobian matrix of the combined model \eqref{eq:bmmodel}, which is formed from the partial derivatives
\begin{equation}
\mathbf{J}_{\mathrm{tree}}[i,j] = \frac{\partial \mathrm{AGB}(\hat{\boldsymbol{\alpha}},d_i,h_i,\mathbf{s}_i) }{\partial \hat{\alpha}_j},
\end{equation}
where $d_i$, $h_i$, and $\mathbf{s}_i$ are the DBH, height and the species of the $i$'th tree.

To produce $\widehat{\mathrm{AGBD}}_{\mathrm{plot}}$, tree-level AGBs of each plot are aggregated and then divided by the plot area. These can be written as matrix operations and thus the covariance of $\widehat{\mathrm{AGBD}}_{\mathrm{plot}}$ is
\begin{equation}
\widehat{\mathbf{C}}_{\mathrm{plot}}=\mathbf{A}^{-1}\mathbf{U}\widehat{\mathbf{C}}_{\mathrm{tree}}\mathbf{U}^T\mathbf{A}^{-1},
\end{equation}
where $\mathbf{A}$ is a diagonal matrix, where $A_{ii}$ is the area in hectares of the $i$'th plot, and $U$ is an aggregation matrix, for which
\begin{equation}
\mathbf{U}_{ij} = \begin{cases} 1, & j\text{'th tree belongs to } i\text{'th plot} \\
0, & \text{otherwise} \end{cases}
\end{equation}

\subsection{Proxy biomass model}
\label{sec:als}
The field plot biomass values $\widehat{\mathbf{AGBD}}_{\mathrm{plot}}$ and metrics from ALS and Sentinel-2 were used to fit a quadratic proxy biomass model with four predictor variables: 
\begin{equation}
\label{eq:alsmodelform}
\mathbf{AGBD}_{\mathrm{proxy}}=\left(\beta_0+\sum_{i=1}^4\beta_i\mathbf{x}^{(i)}\right)^2+\mathbf{e}_f=f(\boldsymbol{\beta},\mathbf{x})+\mathbf{e}_f,
\end{equation}
where $\beta_i$ are the model coefficients, $\mathbf{x}^{(i)}$ the predictors and $\mathbf{e}_f\sim\mathcal{N}\left(0,\boldsymbol{\Sigma}_{f}\right)$ is an additive error term. 

The covariance matrix $\widehat{\mathbf{C}}_{\boldsymbol{\beta}}$ of the model parameters $\boldsymbol{\beta}$ now depends on two sources of uncertainty: 1) the sample used to fit the model \eqref{eq:bmmodel}, and 2) uncertainty of $\widehat{\mathbf{AGBD}}_{\mathrm{plot}}$ used to fit the model \eqref{eq:bmmodel}. Following \citet{saarela2020}, we use the law of total variance and Taylor approximation of the nonlinear model \eqref{eq:bmmodel} to write:
\begin{equation}
\label{eq:alsparcov}
\begin{split}
&\widehat{\mathbf{C}}_{\boldsymbol{\beta}} = (\mathbf{J}_{f}^T\widehat{\boldsymbol{\Sigma}}_{f}^{-1}\mathbf{J}_{f})^{-1}\\
&+ (\mathbf{J}_{f}^T\widehat{\boldsymbol{\Sigma}}_{f}^{-1}\mathbf{J}_{f})^{-1}
\mathbf{J}_{f}^T\widehat{\boldsymbol{\Sigma}}_{f}^{-1} \widehat{\mathbf{C}}_{\mathrm{plot}}\widehat{\boldsymbol{\Sigma}}_{f}^{-1}\mathbf{J}_{f}
(\mathbf{J}_{f}^T\widehat{\boldsymbol{\Sigma}}_{f}^{-1}\mathbf{J}_{f})^{-1},
\end{split}
\end{equation}
where $\mathbf{J}_{f}$ is the Jacobian matrix of the model $f(\boldsymbol{\beta},\mathbf{x})$ with respect to $\beta$, which is formed from the partial derivatives
\begin{equation}
\mathbf{J}_{f}[i,j] = \frac{\partial f(\beta,\mathbf{x}^{(i)})}{\partial \beta_j},
\end{equation}
where $\mathbf{x}^{(i)}$ is the predictor vector of the $i$'th field plot.

The trained proxy AGBD model was then used to predict on the 15 $\times$ 15 m subcells of the ICESat-2 segments and the proxy AGBD for the whole 90 m ICESat-2 segment was acquired by averaging the subcell predictions. By using Taylor approximation and writing the subcell averaging as a matrix operator, the covariance for the 90 m proxy AGBD is
\begin{equation}
\widehat{\mathbf{C}}_{\mathrm{proxy}} = \mathbf{M}^T\mathbf{J}_{f*}^T\widehat{\mathbf{C}}_{\boldsymbol{\beta}}\mathbf{J}_{f*}\mathbf{M},
\end{equation}
where $\mathbf{J}_{f*}$ is the Jacobian matrix of the proxy AGB model evaluated at the 15 m subcell predictor vectors and $\mathbf{M}$ is the averaging matrix, for which
\begin{equation}
\mathbf{M}[i,j] = \begin{cases}
\frac{1}{6}, &\text{subcell $i$ belongs to segment $j$} \\
0, & \text{otherwise.}
\end{cases}
\end{equation}

\subsection{ICESat-2 biomass model}
\label{sec:i2}
The ICESat-2 biomass model had the same form as the proxy biomass model \eqref{eq:alsmodelform}:
\begin{equation}
\label{eq:i2modelform}
\mathbf{AGBD}_{\mathrm{I2}}=\left(\gamma_0+\sum_{i=1}^4\gamma_i\mathbf{y}^{(i)}\right)^2+\mathbf{e}_g=g(\boldsymbol{\gamma},\mathbf{y})+\mathbf{e}_g,
\end{equation}
where $\gamma_i$ are the model coefficients, $\mathbf{y}^{(i)}$ the four ICESat-2 predictors, and $\mathbf{e}_g$ is an additive error term. 

The covariance matrix of the model coefficients $\boldsymbol{\gamma}$ was calculated in a similar way as previously:
\begin{equation}
\begin{split}
&\widehat{\mathbf{C}}_{\boldsymbol{\gamma}} = (\mathbf{J}_{g}^T\widehat{\boldsymbol{\Sigma}}_{g}^{-1}\mathbf{J}_{g})^{-1}\\
&+ (\mathbf{J}_{g}^T\widehat{\boldsymbol{\Sigma}}_{g}^{-1}\mathbf{J}_{g})^{-1}
\mathbf{J}_{g}^T\widehat{\boldsymbol{\Sigma}}_{g}^{-1} \widehat{\mathbf{C}}_{\mathrm{proxy}}\widehat{\boldsymbol{\Sigma}}_{g}^{-1}\mathbf{J}_{g}
(\mathbf{J}_{g}^T\widehat{\boldsymbol{\Sigma}}_{g}^{-1}\mathbf{J}_{g})^{-1},
\end{split}
\end{equation}
where $\mathbf{J}_{g}$ is the Jacobian matrix of the model $g(\boldsymbol{\gamma},\mathbf{y})$ with respect to $\gamma$ evaluated at the Valtimo ICESat-2 segment predictor vectors.

The AGBD values for the Nurmes ICESat-2 segments were then predicted using the fitted ICESat-2 model. The covariance matrix of these predictions is
\begin{equation}
\widehat{\mathbf{C}}_{\mathrm{I2}} = \mathbf{J}_{g*}^T\widehat{\mathbf{C}}_{\boldsymbol{\gamma}}\mathbf{J}_{g*},
\end{equation}
where $\mathbf{J}_{g*}$ is the Jacobian matrix evaluated at the Nurmes ICESat-2 segment predictor vectors.

\subsection{Reference estimate}
\label{sec:ref}
The reference estimate is a hierarchical model-based estimate following \citet{saarela2020} using wall-to-wall predicted AGBD from local ALS and Sentinel-2 data in the Nurmes area.
The model based on the ALS and Sentinel-2 data had the same form as the Valtimo proxy AGBD model (Section \ref{sec:als}). Let us denote the model by $\mathrm{AGBD, ref}=h(\boldsymbol{\delta},\mathbf{z})+\mathbf{e}_\mathrm{ref}$. The covariance of the fitted model parameters $\widehat{\mathbf{C}}_{\boldsymbol{\delta}}$ was then estimated similar to equation \eqref{eq:alsparcov}.

The model based on ALS and Sentinel-2 data was then used to produce a 15 $\times$ 15 m wall-to-wall AGBD raster over the Nurmes area. The hierarchical model-based estimate is then the average of the predicted AGBD. The variance of the average AGBD is
\begin{equation}
\label{eq:hbmcov}
\widehat{\mathrm{Var}}(\hat{\mu}_{\mathrm{ref}})  = \frac{1}{n_{\mathrm{pix}}^2}\boldsymbol{1}^T\mathbf{J}_{h*}^T\widehat{\mathbf{C}}_{\boldsymbol{\delta}}\mathbf{J}_{h*}\boldsymbol{1},
\end{equation}
where $n_{\mathrm{pix}}$ is the total number of forested pixels in the raster.

\end{document}